\documentclass{revtex4}
\usepackage{graphicx}
\usepackage{epsfig}
\usepackage{amsmath}

\newcommand{\beq}{\begin{equation}}
\newcommand{\eeq}{\end{equation}}
\newcommand{\beqd}{\begin{displaymath}}
\newcommand{\eeqd}{\end{displaymath}}
\newcommand{\beqa}{\begin{eqnarray}}
\newcommand{\eeqa}{\end{eqnarray}}

\newcommand{\sign}{{\rm sign}}

\newcommand{\arctanh}{{\rm arctanh}}
\newcommand{\comment}[1]{}

\begin{document}

\title{Diluted Mean-Field Spin-Glass Models at Criticality}

\author{G. Parisi$^1$ F. Ricci-Tersenghi$^1$ and T. Rizzo$^2$} 

\affiliation{
$^1$ Dipartimento di Fisica, IPCF-CNR, UOS Roma, and INFN, Sezione di Roma1,
Universit\`a ``La Sapienza'', Piazzale A. Moro 2, I-00185, Rome, Italy\\
$^2$ IPCF-CNR, UOS Roma, Universit\`a ``La Sapienza'', Piazzale A. Moro 2, I-00185, Rome, Italy\\
}

\begin{abstract}
We present a method derived by cavity arguments to compute the spin-glass and higher-order susceptibilities in diluted mean-field spin-glass models.
The divergence of the spin-glass susceptibility is associated to the existence of a non-zero solution of a homogeneous linear integral equation. 
Higher order susceptibilities, relevant for critical dynamics through the parameter exponent $\lambda$, can be expressed at criticality as integrals involving the critical eigenvector. The numerical evaluation of the corresponding analytic expressions is discussed. The method is illustrated in the context of the de Almeida-Thouless line for a spin-glass on a Bethe lattice but can be generalized straightforwardly to more complex situations. 
\end{abstract}

\maketitle

\section{Introduction}

In disordered magnetic systems the spin-glass (SG) singularity occurs by definition for those values of the external parameters where the spin-glass susceptibility diverges \cite{MPV}. The computation of this four-point correlation function in the paramagnetic phase is therefore tantamount to the location of the phase transition.
Higher order (six point) susceptibilities also play an important role because they determine quantitatively the function $q(x)$ in the Replica-symmetry-breaking phase in the vicinity of the critical point \cite{Gross85,Parisi13}. Recently it has been discovered that they also determine quantitatively the non-universal dynamical critical exponents \cite{Parisi13,calta1,calta2,calta3,ferra1}.
Furthermore the same equilibrium susceptibilities are  important for off-equilibrium behavior \cite{Caltagirone13,Rizzo13}. In this paper we discuss the problem of the computation of these susceptibilities in mean-field spin-glass models with finite connectivity. 

In mean-field spin-glass models, both fully-connected or with finite connectivity, one can  use the replica-method in order to write down a saddle-point expression for the free energy and determine the location of the phase transition in parameter space by studying the stability of the paramagnetic solution. In fully-connected models, like the Sherrington-Kirkpatrick (SK) model, the order parameter is a $n \times n$ matrix and this program can be completed both in the paramagnetic and in the spin-glass phase \cite{MPV}. 
In the case of models with finite connectivity the replicated order parameter is a more complicated object and the computations are more difficult \cite{DeDominicis89}, on the other hand one can exploit the (local) tree-like structure of the corresponding graphs and apply instead the cavity method, thus avoiding replicas \cite{Mezard01}.

By means of the cavity method it is rather easy to obtain a self-consistent equation for the order parameter which, in the paramagnetic phase is a probability density of the local cavity fields. However the self-consistent equation and its solution are perfectly regular at the SG transition and cannot be used to locate it (except in the case of strictly zero external field). Previous studies in the context of the replica method has shown that the critical point is associated instead to the solution of certain integral equations \cite{Weigt96,Janzen10b} and the same equations have also been rederived in the context of the cavity method \cite{Janzen10a}.  
The cavity method derivation relies essentially on joint iterative equations for the fields and the susceptibilities, a technique that have been developed originally for the study of the number of metastable states on locally tree-like models \cite{Parisi05}. The derivation allows also to understand the connection between the integral equations and numerical methods based on coupled systems that allowed the first quantitative description of the region of validity of the paramagnetic phase \cite{PPR}.
In this paper we present an alternative cavity method derivation of these integral equations and discuss their numerical solution down to zero temperature. This discussion is instrumental to the main new result that we report here: {\it i.e.} the expression, derived by cavity arguments, of the two {\it static} six-point susceptibilities that control critical {\it dynamics}.

We will illustrate the method in the context of the de Almeida-Thouless (dAT) transition on a Ising SG defined on a random lattice with fixed connectivity, but it can be generalized straightforwardly to more complicated models in order to obtain the corresponding expressions for the same six-point susceptibilities. These extensions include {\it e.g.} Potts spins, fluctuating connectivity, $p$-spin interactions. It can also be applied to different kind of SG phase transitions including notably some instances of discontinuous Replica-Symmetry-Breaking transitions that display the phenomenology of structural glasses.

The plan of the paper is as follows. 
In the next section  we will present the results in a concise way together with their physical motivations.  In section \ref{derbethe} we will derive the integral equation condition and we will discuss its numerical solution down to zero temperature. 
In section \ref{cubic} we will present the derivation of the six-point susceptibilities and use it to determine them on the dAT line in the case of a SG model with connectivity $c=4$. 
In section \ref{Conclusions} we give our conclusions. In the appendix we report the detailed analysis of the high-connectivity (SK) limit.

\section{Outline of the Results}
\label{outline}

The spin-glass transition is characterized by the divergence of the spin-glass susceptibility $\chi_{SG}$ defined as:
\beq
\chi_{SG}\equiv{1\over N}\sum_{ij}\overline{(\langle s_is_j \rangle-\langle s_i\rangle \langle s_j \rangle)^2}
\label{chiSG}
\eeq
where the angular brackets mean thermal average and the overline mean disorder average. 
Dynamics is also critical at the phase transition. In particular the time decay of the correlation $C(t)\equiv N^{-1}\sum_i^N \overline{\langle s_i(0)s_i(t) \rangle}$ is exponential in the paramagnetic phase but becomes power-law at the critical point:
\beq
C(t) \simeq q_{EA}+{c \over t^a} 
\eeq
where $q_{EA}$ is the Edwards-Anderson parameter.
It has been recently established \cite{Parisi13} that
the dynamical exponent $a$ can be computed from the ratio of two {\it static} six-point susceptibilities, more precisely we have:
\beq
{\Gamma^2(1-a) \over \Gamma(1-2 a)}={\omega_2 \over \omega_1}
\label{uno}
\eeq
where $\Gamma(x)$ is the Gamma function and $\omega_2$, $\omega_1$ are defined as:
\beq
\omega_1 \equiv {1 \over N} \sum_{ijk}\overline{\langle s_i s_j\rangle_c \langle s_j s_k\rangle_c \langle s_k s_i\rangle_c}
\label{defome1}
\eeq
\beq
\omega_2 \equiv {1 \over 2 N} \sum_{ijk}\overline{\langle s_i s_j s_k\rangle_c^2 }
\label{defome2}
\eeq
where the suffix $c$ means connected correlations \cite{Parisi13}.
Note that in the literature it is often introduced the so-called parameter exponent $\lambda$ that controls $a$ through $\Gamma^2(1-a)/ \Gamma(1-2 a)=\lambda$, in terms of $\lambda$ eq. (\ref{uno}) reads $\lambda=\omega_2/\omega_1$.
Besides these more recent developments it has been known \cite{Gross85,MPV,Rizzo13} that the very same ratio $\omega_2/\omega_1$ is equal to the position of the breaking point in continous RSB transitions. For instance in the RSB phase near the dAT line (that will be studied in the following) this ratio is precisely equal to the point $x$ where the function $q(x)$ displays a continuous part. We will provide a general method to obtain the expressions of $\chi_{SG}$, $\omega_1$ and $\omega_2$ in models with finite connectivity. 

In finite connectivity models the paramagnetic phase can be described through a self-consistent equation for the distribution of the fields.
In the following we specialize to the case of Ising spins in presence of a field $H$ interacting by means of two-body quenched couplings $J_{ij}$ on a random regular graph {\it i.e.}  a random  graph with fixed connectivity $c=M+1$. In the following, with a slight abuse of notation, we will also refer to this kind of graph as a Bethe lattice. The  relevant iterative equation is \cite{Mezard01}:
\beq
P(u)=\int P_M(u_M)du_M\; \overline{\delta\Big(u-\tilde{u}(J,u_M+H)\Big)}
\label{PU}
\eeq 
with the overline being the average with respect to the distribution of the quenched coupling $J$ and
\beq
\tilde{u}(J,h) \equiv {1 \over \beta}\arctanh(\tanh\beta J \tanh \beta h) \ .
\eeq
The function $P_{M}$ is the distribution of the sum of $M$ independent fields, each one distributed according to $P$, i.e.,
\beq
P_K(u)=\int \prod_{i=1}^K P(u_i)\,du_i\;\delta\left(u-\sum_{i=1}^K u_i\right) \ .
\label{PUM}
\eeq
We will show that the dAT line, where by definition $\chi_{SG}$ diverges, is specified by the condition that the following homogeneous linear equation admits a non-zero solution $g(u)$:
\beq
g(u) = M \int du_M\,du_1\,P_{M-1}(u_M-u_1)g(u_1)\;
\overline{ \delta\Big(u-\tilde{u}(J,u_M+H)\Big)
\left( {d \tilde{u} (J,u_M+H) \over d H } \right)^2 }
\label{DATBETHE}
\eeq
where the derivative inside the integral reads:
\beq
 {d \tilde{u} (J,h) \over d h }= \frac{\tanh(\beta J) [1-\tanh(\beta h)^2]}{[1-\tanh(\beta J)^2\tanh(\beta h)^2]}
\eeq
Then we will show that the six-point susceptibilities needed to determine the parameter exponent at criticality can be expressed in term of the eigenvector $g(u)$ of the integral equation (\ref{DATBETHE}). More precisely one obtains:
\beq
{\omega_2 \over \omega_1}={ \langle\langle 2 m_0^2(1 - m_0^2)^2 \rangle\rangle \over \langle\langle (1 - m_0^2)^3 \rangle\rangle}
\label{w2suw1}
\eeq
where 
\beq
m_0= \tanh \beta[u_1+u_2+u_3+u_{M-2}+H]
\label{u0}
\eeq
and
\beq
\langle\langle \cdots \rangle\rangle \equiv \int du_1 du_2 du_3 du_{M-2} \, g(u_1)g(u_2)g(u_3)P_{M-2}(u_{M-2}) \cdots \ .
\eeq
Note that since eq. (\ref{DATBETHE}) is homogeneous the eigenvector $g(u)$ is specified up to a normalization constant but the ratio $\omega_2/\omega_1$ is independent of it.

As discussed in \cite{calta1,Parisi13} the connection between that parameter exponent $\lambda$ and the ratio $\omega_2/\omega_1$ is rather general and holds not only for the SG transition in a field by also in the case of discontinuous SG transitions described dynamically bu the Mode-Coupling-Theory phenomenology. Furthermore it has been shown that the ratio $\omega_2/\omega_1$ plays also a crucial role in off-equilibrium dynamics \cite{Caltagirone13,Rizzo13}.  In order to realize these different types of transitions one can consider for instance SG models with $p$-spin interactions or with Potts spins. Although in this paper we shall only consider the case of Ising spins with two-body interactions on a fixed-connectivity graphs, we stress once again that analogous expressions can be obtained in more complex situations through straightforward extensions of the cavity arguments used in the following.

We note that the expression of the susceptibility can be also generalized, indeed the above equation for the critical condition is an instance of a sequence of eigenvalue equations of the general form
\beq
\mu_k g(u) =  \int du_M\,du_1\,P_{M-1}(u_M-u_1)g(u_1)\;
\overline{ \delta\Big(u-\tilde{u}(J,u_M+H)\Big)
\left( {d \tilde{u} (J,u_M+H) \over d H } \right)^k }
\label{DATBETHEK}
\eeq
that can be used in order to obtain higher order moments of the susceptibility, see \cite{RFIM_LD} where this method has been applied in order to study the multi-fractal distribution of connected correlations at large distance. 

\section{The equation for the critical point}
\label{derbethe}

\subsection{Derivation of the equation}

A derivation of the condition (\ref{DATBETHE}) by means of the cavity method has been given in \cite{Janzen10a}. In this section we will present an alternative derivation which is the key to unveil the connection between the critical eigenvector and the computation of the six-point susceptibilities (which are also related to cubic cumulants of the order parameter). Our starting point is the spin-glass susceptibility that, due to the average over disorder, can be rewritten with respect to a given site $s_0$ of the Bethe lattice as:
\beq
\chi_{SG}=\sum_{i}\overline{\langle s_0s_i \rangle_c^2}=\sum_i \overline{ \left( dm_0 \over dH_i \right)^2}
\eeq
where $m_0$ is the magnetization of the root $s_0$ and $H_i$ is a local field on site $i$.
For a given site $i$ we define its father $j=F(i)$ as the spin $j\in \partial 0$, with $\partial 0$ being the set of neighbors of 0, such that $i$ is connected to $0$ through $j$.
On the other hand the magnetization on the root can be written as:
\beq
m_0=\tanh \beta h_0 \ \ ,  h_0=H_0+\sum_{j\in \partial 0} u_{j \rightarrow 0} 
\eeq
where $u_{j \rightarrow 0}$ is by definition the field acting on site zero when all its neighbors except $j$ are removed (in the language of computer science it would be the message passed from site $j$ to site zero). 
Therefore we have:
\beq
{dm_0 \over dH_i}=(1-m_0^2) {du_{j \rightarrow 0} \over dH_i}  \ \  j=F(i)  
\eeq
Due to the locally tree-like nature of the lattice the field $u_{j \rightarrow 0}$ is influenced only by a field on one of its sons $i \in S(j)$ defined such that $j=F(i)$ therefore we may write:
\beq
\sum_i  \left( dm_0 \over dH_i \right)^2 = (1-m_0^2)\left[ 1+\sum_{j \in \partial 0} \, \sum_{k\in S(j)}  \left({du_{j \rightarrow 0} \over dH_k}\right)^2\right]
\label{support}
\eeq
where in the above expression the $1$ is present in order to take into account of the case in which the site $i$ is the root itself.    
At this point we introduce the following physical object in order to average over the disorder:
\beq
\chi(u) \equiv \overline{\delta(u-u_{j \rightarrow 0}) \sum_{k\in S(j)}  \left({du_{j \rightarrow 0} \over dH_k}\right)^2  }
\label{guphys}
\eeq
In principle we should have written $\chi_j(u)$ but the difference between different branches has disappeared due to the disorder average.
In physical terms $\chi(u)$ is essentially the Spin-Glass susceptibility of a given branch conditioned to the fact that the value of the field $u_{j \rightarrow 0}$ is $u$. Indeed using eq. (\ref{support}) we can see that the total $\chi_{SG}$ can now be written as an integral of $\chi(u)$ over possible values of $u$:
\beqa
\chi_{SG} & = & \int P_{M+1}(u) [1-\tanh^2(\beta H+ \beta u)]^2\,du  \ +
\nonumber
\\
& + & (M+1)\int P_{M}(u')\chi(u'') [1-\tanh^2(\beta H+ \beta (u'+u''))]^2\, du' du''
\eeqa
Performing essentially the same steps as for the total $\chi_{SG}$ one can obtain the following iterative equation for the function $\chi(u)$:
\beqa
\chi(u) & = & \int P_M(u')\, \overline{\delta[u-\tilde{u}(J,u'+H)]\left(d\tilde{u}(J,u'+H) \over dH\right)^2} du'+
\nonumber
\\
& + & M \int P_{M-1}(u')\chi(u'')\overline{\delta[u-\tilde{u}(J,u'+u''+H)]\left(d\tilde{u}(J,u'+u''+H) \over dH\right)^2}du'du''
\eeqa
where we have used the definitions of the previous section. Note that we need the whole function $\chi(u)$ in order to write the iterative equation and this why we introduced it in the first place. The above equation can be solved leading to a finite $\chi(u)$ and $\chi_{SG}$ provided the linear system is invertible. This is not possible, meaning that we are at a critical point, if the corresponding homogeneous linear system {\it i.e.} eq. (\ref{DATBETHE}) admits a non-zero solution 
thus completing our argument. 
The function $\chi(u)$ diverges at the critical point and standard arguments tell us that the critical eigenvector $g(u)$ controls its divergence, more precisely we have:
\beq
\chi(u) \propto {g(u) \over \tau} 
\label{chig}
\eeq
where $\tau$ depends on the external parameters ({\it e.g.} temperature and field) and vanishes linearly at the critical point.

\subsection{Solving the critical equation}

Now we want to show how to actually solve Eq.~(\ref{DATBETHE}) and to connect it to the original method for computing the dAT line.
The standard way to compute $P(u)$ from Eq.~(\ref{PU}) is by population dynamics: the function $P(u)$ is approximated by a population of $N$ fields, $P(u) = N^{-1} \sum_{i=1}^N \delta(u-u_i)$, that plugged on the rhs produces a new sum of delta functions, that is a new population. Iterating this process several times the population may converge to a good approximation for the $P(u)$ that solves the self-consistency equation (\ref{PU})

The computation of $g(u)$ from Eq.~(\ref{DATBETHE}) is not straightforward.
Indeed, if both $P(u)$ and $g(u)$ are approximated by populations, then the rhs of Eq.~(\ref{DATBETHE}) would result in a {\em weighted} population, due to the extra factor
\[
f(\beta J, \beta u) = \left(\frac{\tanh(\beta J) [1-\tanh(\beta h)^2]}{[1-\tanh(\beta J)^2\tanh(\beta h)^2]}\right)^2
\]
Working with a weighted population is not a good idea, because if the weights becomes very different, then the effective size of the population gets reduced: just to illustrate the concept with an extremal case, if half of the population elements gets a null weight the effective size of the population gets reduced by at least a factor 2.

The problem of solving a self consistent integral equation containing a reweighting term $f(\beta J, \beta u)$ is not new, as it appears e.g. in 1RSB equations obtained by the replica method \cite{PRL2001} or the cavity method \cite{Mezard01} and even in more complicated equation obtained by the replica cluster variational method \cite{RCVM}.

A possible way to solve these equation is that of discretizing the $g(u)$ by approximating it with a histogram of $N$ bins. The fact that Eq.(3) is linear in $g(u)$ implies that the equations for the $N$ heights of the histogram bins are again linear. In practice one should compute the largest eigenvalue of a random $N\times N$ matrix that depends on the fixed point $P(u)$ (which can be kept as a population): when this eigenvalue equals 1 then Eq.~(\ref{DATBETHE}) is satisfied and the system is at the critical point.

We prefer to approximate $g(u)$ by a population (as we always do for $P(u)$ as well) and we devise two different methods for solving Eq.~(\ref{DATBETHE}).

In the first method, the factor $f(\beta J, \beta u)$ is interpreted as a the probability that the newly generated element should be included in the new population representing $g(u)$. In the present case we have that $0 \le f(\beta J, \beta u) \le 1$ and so the interpretation as a probability is straightforward. In more complicated cases \cite{RFIM_LD} the reweighting factor may be larger than 1 and in that case more than one copy of the same new element should be eventually included in the new population. If this is the case, we suggest to make the new population larger than the old one, and then filter it by randomly choosing its elements: in this way a much smaller fraction of twin element will finally remain in the new population and the information content of the population is preserved.

The second method is essentially equivalent to the original method invented to identify the location of the dAT line in sparse models \cite{PPR}. Each cavity field $u_i$ is perturbed by an infinitesimal quantity $\delta u_i$ and the evolution of the pairs $(u_i,\delta u_i)$ is followed according to the BP equations. Thanks to the symmetry of the interactions, we have that $\langle\delta u_i|u_i=u\rangle=0$ for any $u$ value and the interesting quantities to look at are the variances, that evolve under BP by the following equation
\begin{equation}
\langle \delta u^2 | u \rangle_{t+1} = M \int du_1 \langle \delta u^2 | u_1 \rangle_t \prod_{i=2}^M dP(u_i)\;\overline{ \delta\left(u-\tilde{u}\Big(J,H+\sum_{i=1}^M u_i\Big)\right)
\left( \frac{\tanh(\beta J) [1-\tanh(\beta h)^2]}{[1-\tanh(\beta J)^2\tanh(\beta h)^2]} \right)^2 }
\label{eq:d1}
\end{equation}
that corresponds to Eq.~(\ref{DATBETHE}) by equating $g(u)=\langle\delta u_i^2 | u_i=u\rangle$ in the large time limit.
Eq.~(\ref{eq:d1}) has a non zero solution only at the critical point. So in order to measure $g(u)=\langle\delta u^2|u\rangle$ also away from the critical point one can renormalize it at each BP step and this corresponds to solve the following equation
\begin{equation}
g(u) = \mu M \int du_1 g(u_1) \prod_{i=2}^M dP(u_i)\;\overline{ \delta\left(u-\tilde{u}\Big(J,H+\sum_{i=1}^M u_i\Big)\right)
\left( \frac{\tanh(\beta J) [1-\tanh(\beta h)^2]}{[1-\tanh(\beta J)^2\tanh(\beta h)^2]} \right)^2 }
\label{eq:d2}
\end{equation}
where $\mu$ is the inverse of the normalization factor in the large time limit. The above equation no longer depend of time, but it only involves asymptotic quantities and the new parameter $\mu$. It admits a non-zero solution at any temperature and external field.
Interpretation of Eq.(\ref{eq:d2}) is straightforward: in the high temperature paramagnetic phase $\mu < 1$, so any perturbation goes to zero exponentially as $\mu^t$ and the BP fixed point is stable; in the low temperature spin glass phase $\mu > 1$, a perturbation grows as $\mu^t$ and the BP fixed point is unstable (indeed the correct solution is provided by an Ansatz breaking the replica symmetry).

In practice, after having computed the $P(u)$ from Eq.~(\ref{PU}) by population dynamics, we solve Eq.~(\ref{eq:d2}), by one of the two methods described above, and we compute the maximum eigenvalue $\mu$ of the integral kernel and the corresponding eigenvector $g(u)$.

\begin{figure}
\includegraphics[width=0.49\columnwidth]{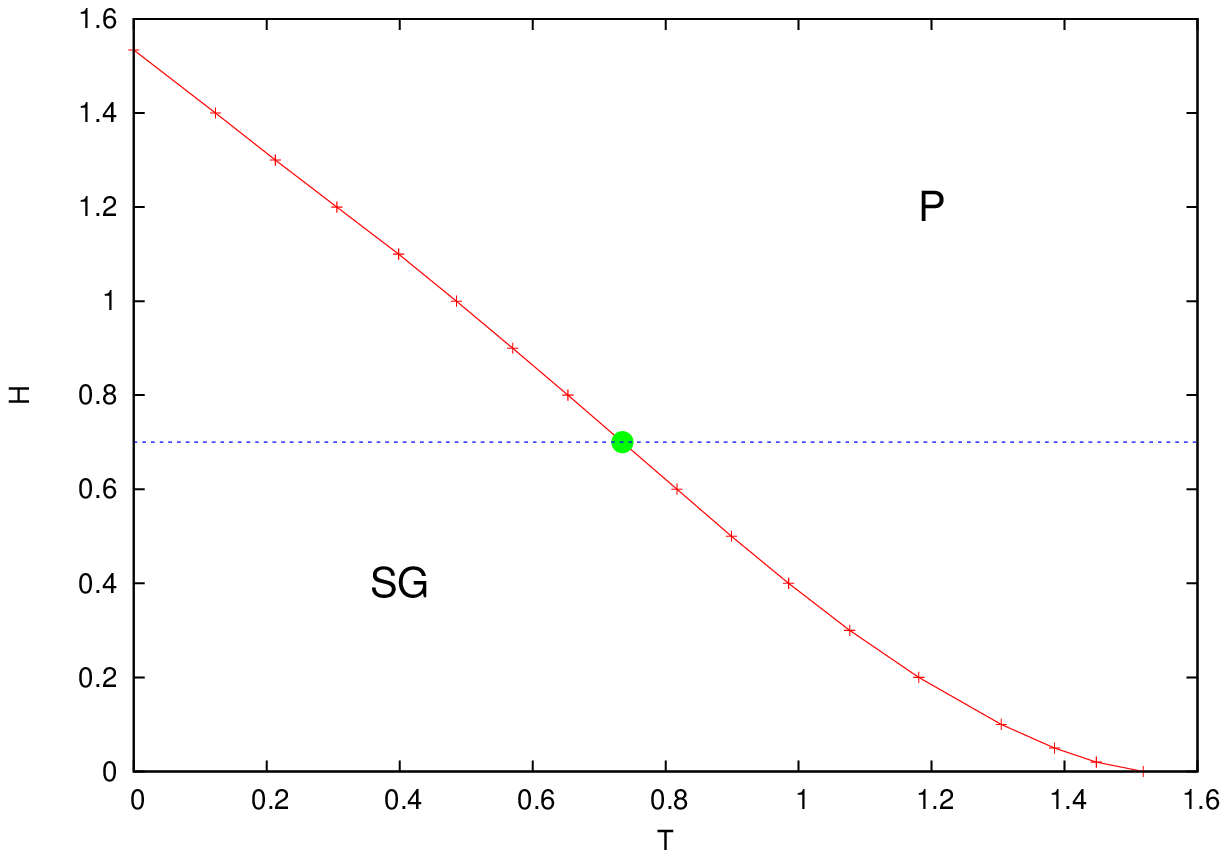}
\includegraphics[width=0.5\columnwidth]{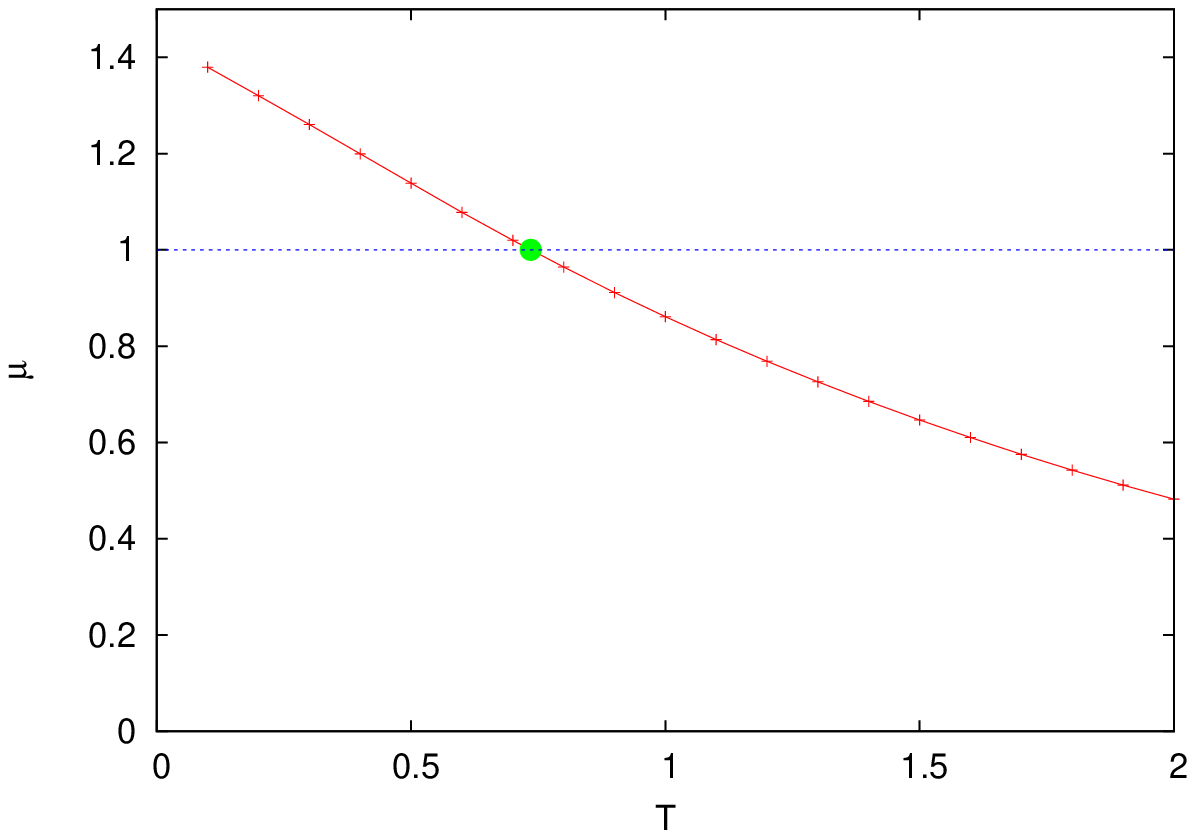}
\caption{Left panel: Critical dAT line for a spin glass model with couplings $J_{ij}=\pm1$ and external field $h$ on a random regular graph (Bethe lattice) of fixed degree $M+1=4$.
Right panel: Maximum eigenvalue $\mu$ of the integral kernel in Eq.~(\ref{eq:d2}) computed along the blue line in the left panel ($H=0.7$).}
\label{fig:dATline}
\end{figure}

\begin{figure}
\includegraphics[width=0.55\columnwidth]{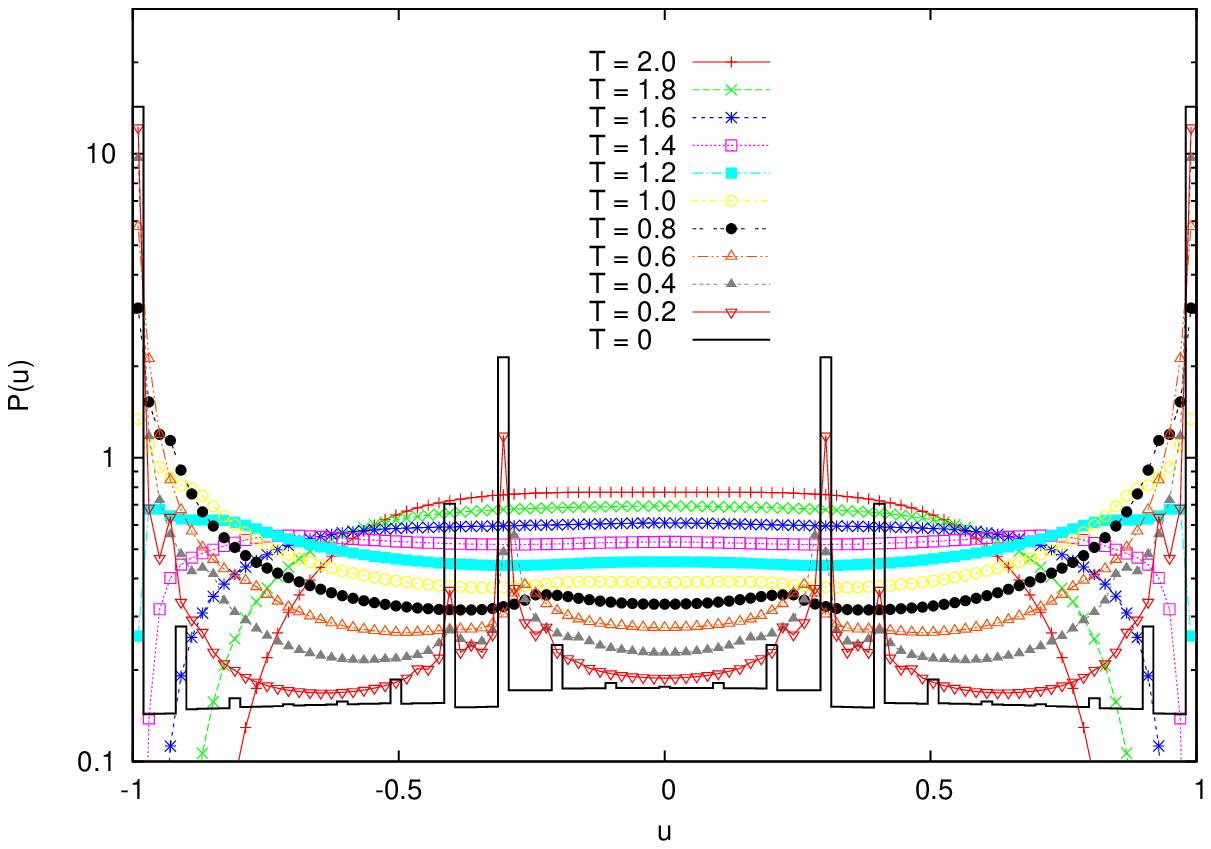}
\includegraphics[width=0.55\columnwidth]{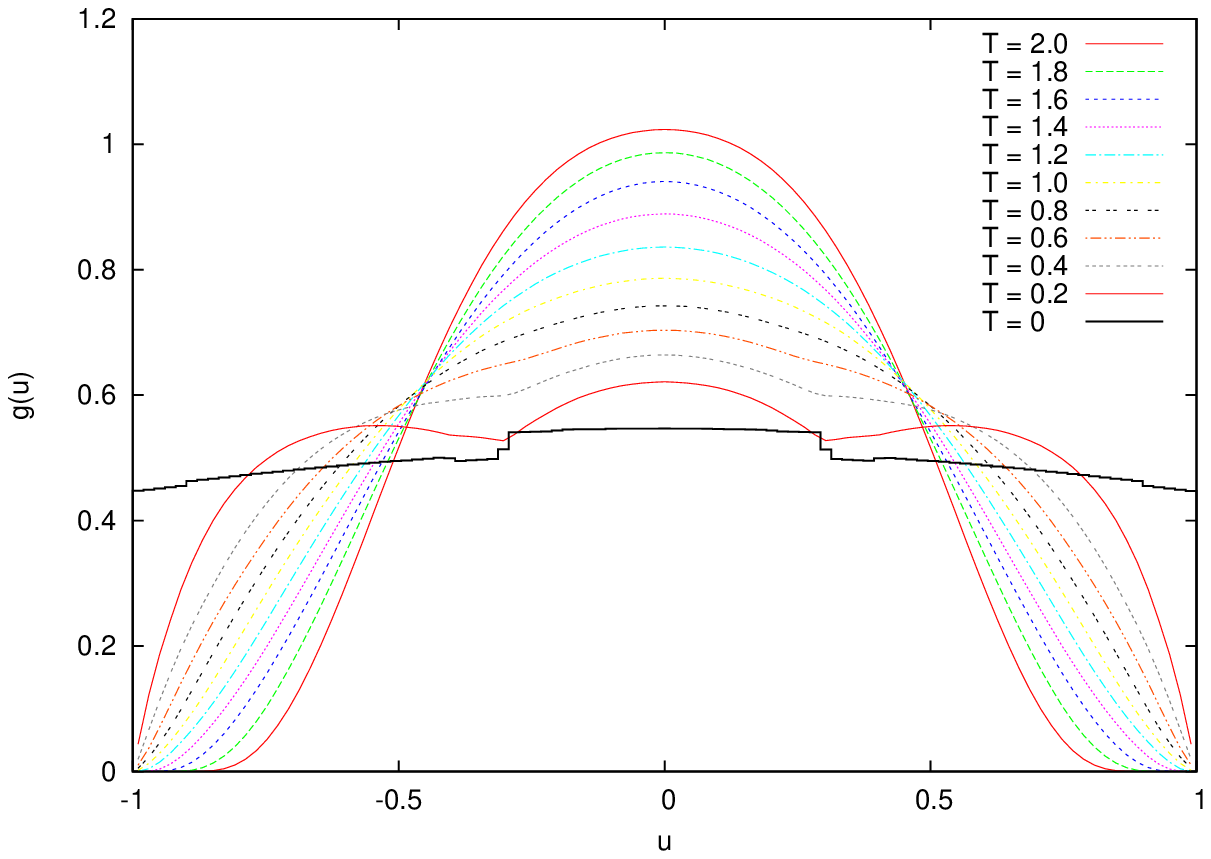}
\caption{Fixed point $P(u)$ (upper panel) and $g(u)$ (lower panel) for $H=0.7$ and several temperatures (including $T=0$).}
\label{fig:Pu}
\end{figure}

We present data obtained for a spin glass model ($J_{ij}=\pm1$ with equal probabilities and uniform external field $h$) on a random regular graph (Bethe lattice) with fixed degree $M+1=4$. The dAT line for this model was already presented in \cite{PhilMag2012} and is reproduced in Fig.~\ref{fig:dATline} (left panel) for readability.
In Fig.~\ref{fig:dATline} (right panel) we show the maximum eigenvalue $\mu$ as a function of the temperature at a fixed field $H=0.7$ (horizontal line in the left panel): the behavior is exactly the one discussed above.

In Fig.~\ref{fig:Pu} (upper panel) we show the fixed point distribution of cavity fields, $P(u)$, at several temperatures and fixed external field $H=0.7$ (please note that the $y$ axis is in log scale). It is worth noticing that the $P(u)$ becomes broader by lowering the temperature, but has no particular change at the critical temperature, $T_c(H=0.7)=0.7353$, and finally becomes singular at zero temperature (we comment more on this below).
In Fig.~\ref{fig:Pu} (lower panel) we show the eigenfunction $g(u)$ corresponding to the maximum eigenvalue $\mu$. It is worth noticing that these functions are even smoother than the corresponding $P(u)$ and even in the $T=0$ limit $g(u)$ remain continuous, although with steps (further comments below).

\begin{figure}
\includegraphics[width=0.55\columnwidth]{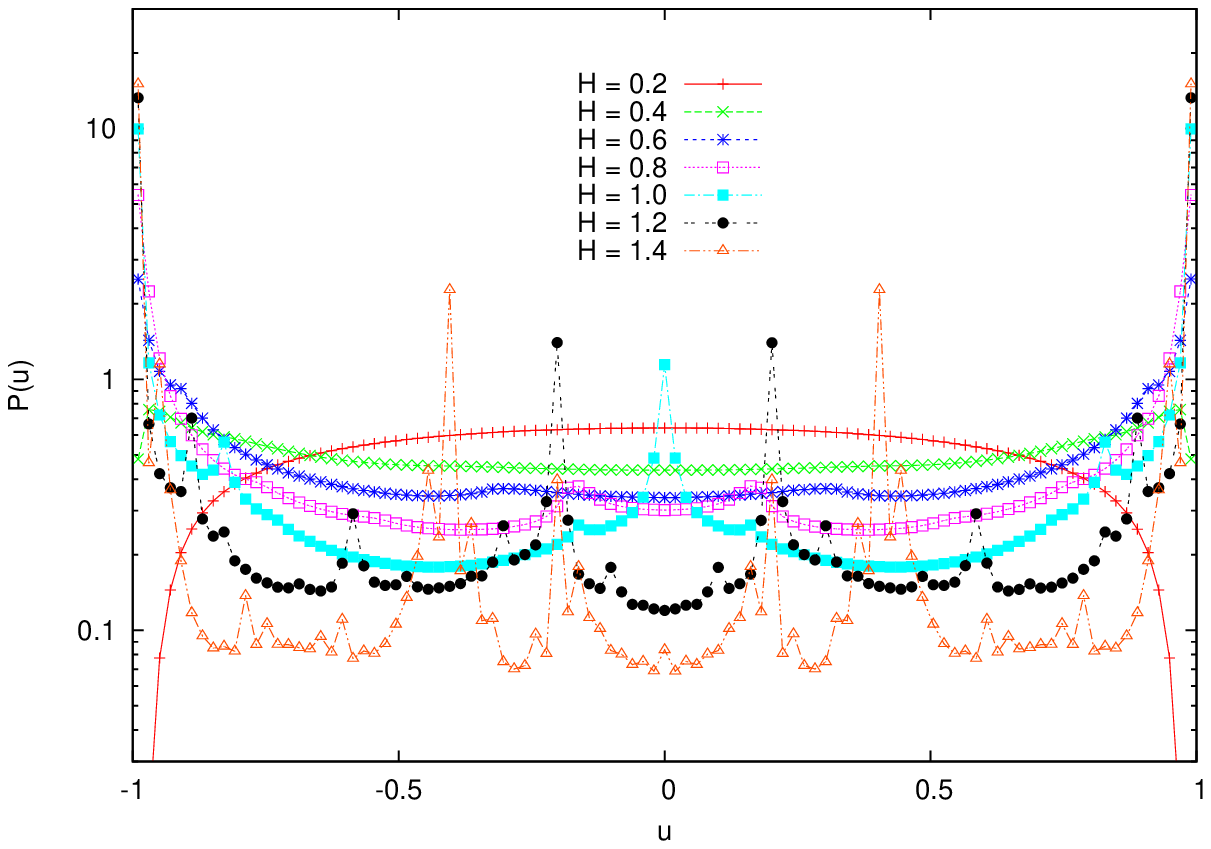}
\includegraphics[width=0.55\columnwidth]{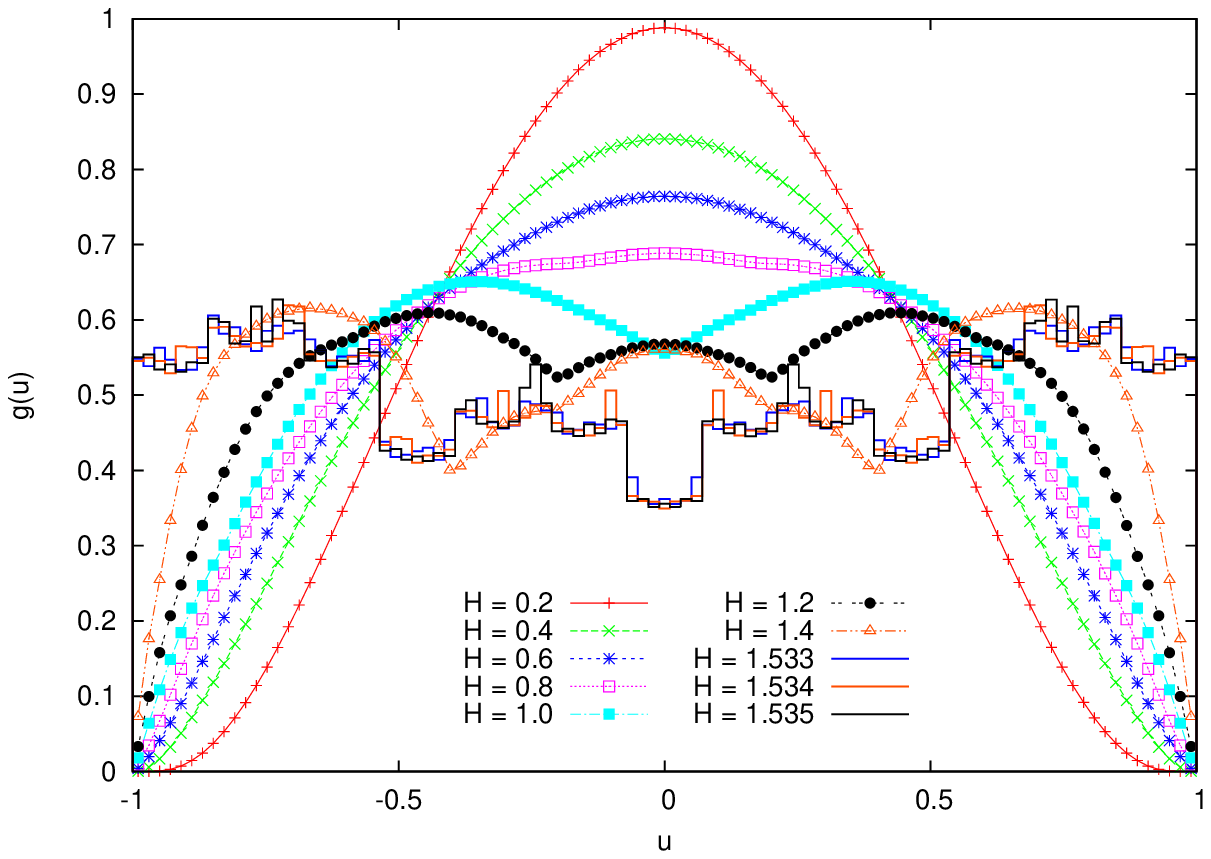}
\caption{Fixed point $P(u)$ (upper panel) and $g(u)$ (lower panel) for several points along the critical dAT line, indexed by the corresponding field value. The lower panel also shows the $g(u)$ computed at $T=0$ with 3 field values, all compatible with the our best estimate for $H_c=1.534(1)$.}
\label{fig:Pu_Tc}
\end{figure}

The method presented in this manuscript is perfectly suitable for studying critical properties of disordered models defined on random graphs: indeed the functions $P(u)$ and $g(u)$ are well defined on the entire critical line and smooth enough (infinitely differentiable) for any $T>0$. Even at $T=0$ they are well defined distributions, that leads to smooth physical observables, once integrated over.

In Fig.~\ref{fig:Pu_Tc} we show these functions computed at several points along the critical line, including the $T=0$ critical point for $g(u)$. Actually in the lower panel of Fig.~\ref{fig:Pu_Tc} we have included three different $g(u)$ computed at $T=0$ with field values which are all compatible with our best estimate for the critical field, $H_c=1.534(1)$.
The comparison of these three distributions should make the reader aware of which features of the critical $g(u)$ at $T=0$ are robust with respect to very small field fluctuations and which are not.

Once we have under control the process for computing the critical distributions $P(u)$ and $g(u)$ along the entire critical line, we can use the resulting data to estimate universal quantities of physical interest.

\begin{figure}
\includegraphics[width=0.55\columnwidth]{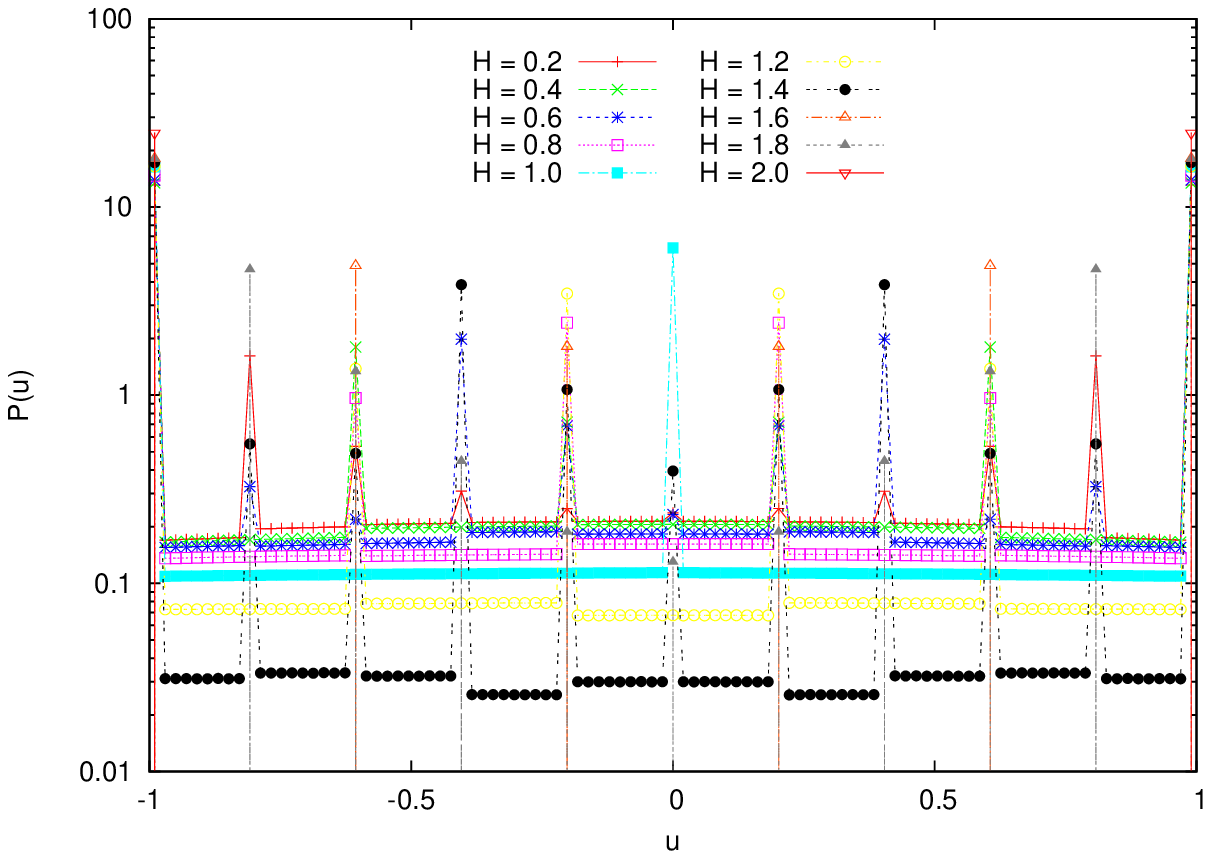}
\includegraphics[width=0.55\columnwidth]{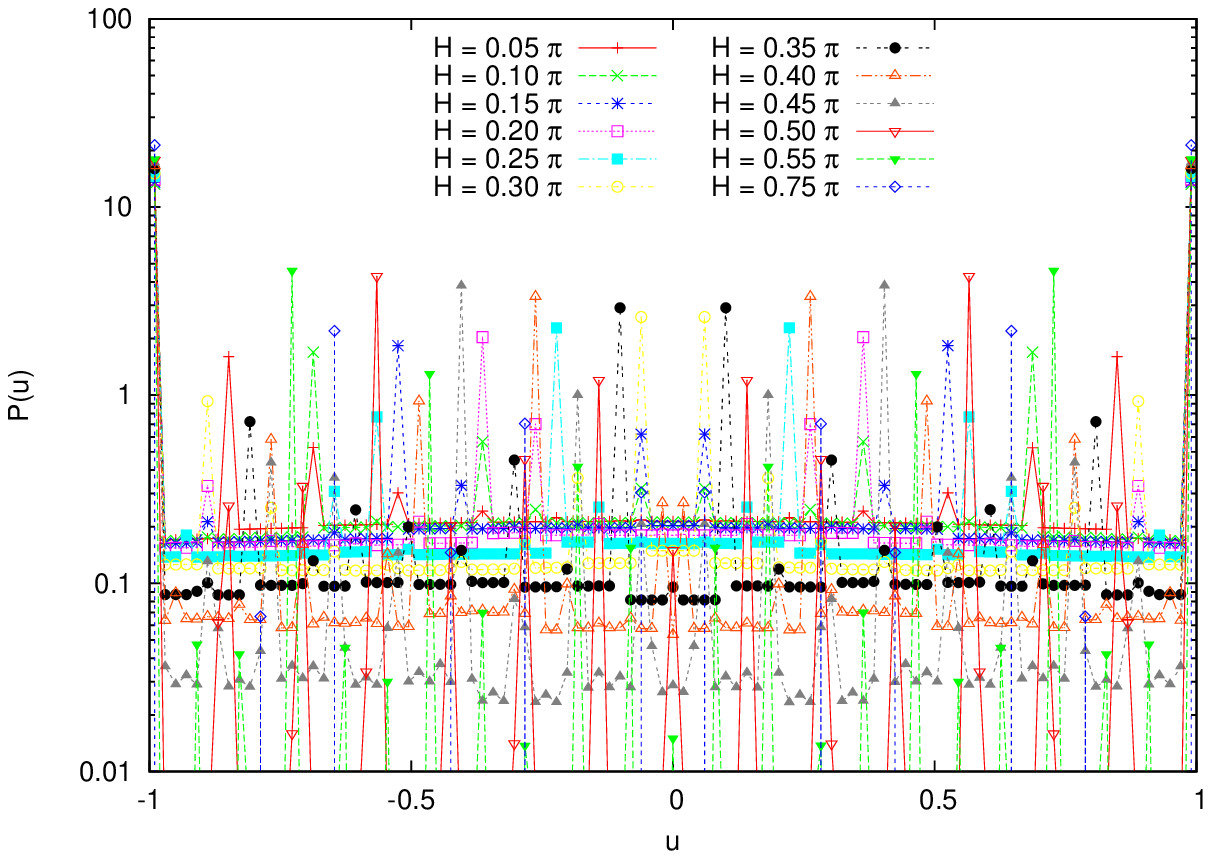}
\caption{Fixed point $P(u)$ for $T=0$ and several external fields.}
\label{fig:Pu_T0}
\end{figure}

\subsection{The zero temperature limit}
\label{T0limit}

The computation of functions $P(u)$ and $g(u)$ at $T=0$ requires some more care, because these functions may develop singularities.
The BP equation to be satisfied by the cavity fields population $P(u)$ is the following
\begin{equation}
P(u) = \int \prod_{i=1}^M P(u_i)\;\overline{\delta\left(u-\hat{u}_J(H + \sum_i u_i)\right)}
\end{equation}
with $\hat{u}_J(x) = \sign(Jx) \min(|x|,1)$, where we have assumed $|J|=1$ without loss of generality.
The function $\hat{u}$ essentially moves the weight of fields such that $|H + \sum_i u_i|>1$ on the extrema of the allowed domain $u \in [-1,1]$. So the fixed point function $P(u)$ is a distribution with at least two delta functions in $u=1$ and $u=-1$. Depending on the value of the external field $H$, further delta peaks are present in $P(u)$ on values $u=n|J|+mH$ with integer valued $n$ and $m$.

In Fig.~\ref{fig:Pu_T0} (upper panel) we show distributions $P(u)$ computed at $T=0$ with $H$ being multiple of $\Delta=0.2$ and the presence of peaks equally spaced by $\Delta$ is evident. Such a regularity in peaks location is present only if the external field and the coupling interaction can be written as $H=n_1 \Delta$ and $|J|=n_2 \Delta$, with integer valued $n_1$ and $n_2$, and $\Delta$ being the peak distance.
For example in Fig.~\ref{fig:Pu_T0} (lower panel) we show distributions $P(u)$ computed with an external field that does not satisfies the above requirement and indeed peaks have less regular positions.

\begin{figure}
\includegraphics[width=0.55\columnwidth]{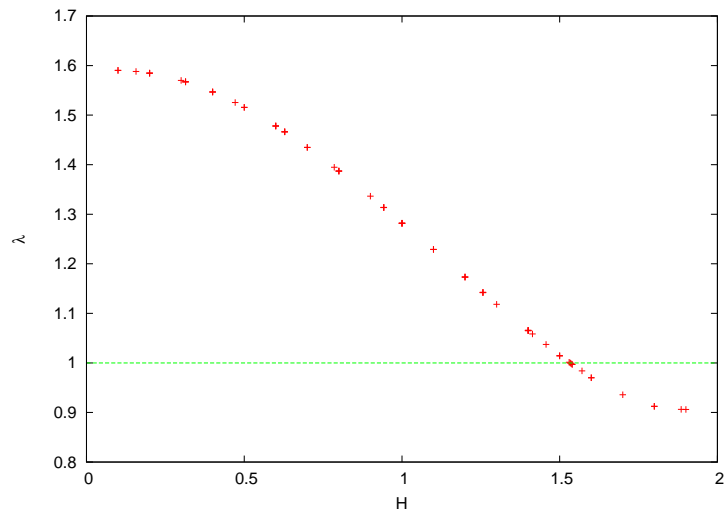}
\caption{Maximum eigenvalue $\mu$ of the $T=0$ integral kernel in Eq.~(\ref{DATBETHET0}) as a function of the external field $H$.}
\label{fig:lambda_h}
\end{figure}

\begin{figure}
\includegraphics[width=0.55\columnwidth]{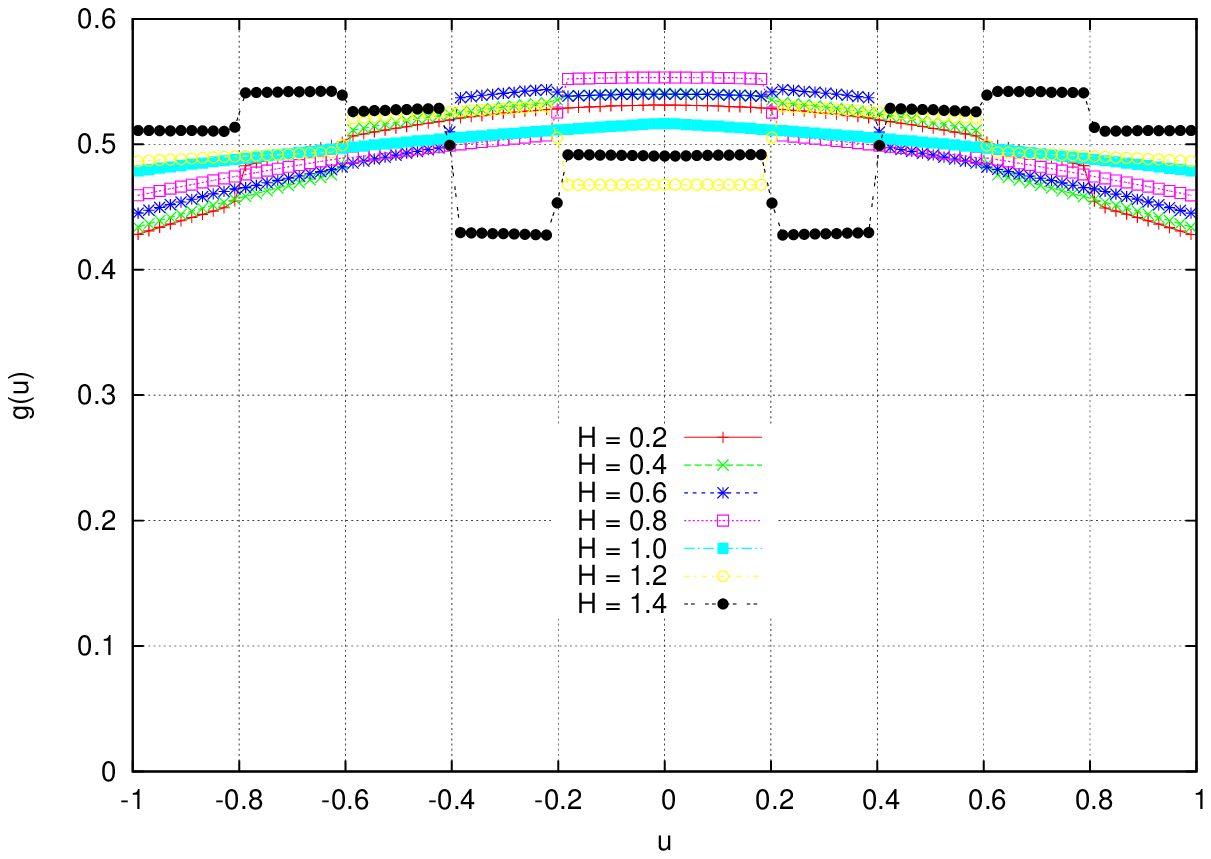}
\includegraphics[width=0.55\columnwidth]{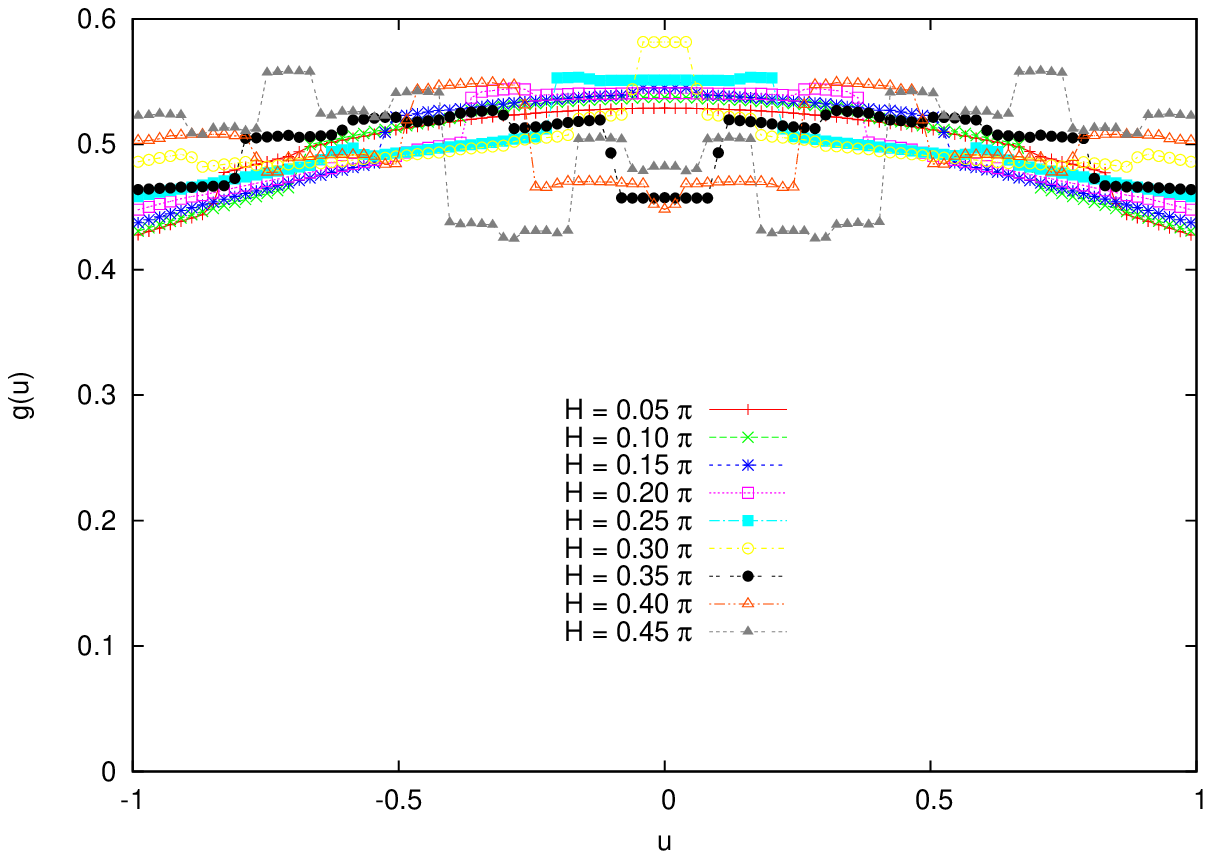}
\caption{Fixed point $g(u)$ for $T=0$ and several external fields.}
\label{fig:gP_T0}
\end{figure}

What is more interesting to notice is the continuous part between the peaks: this ``background'' only exists in the low temperature spin glass phase where the replica symmetry should be broken, as it was already noticed in Ref.~\cite{MPR_JPA04}. The reason for this is simple: in the paramagnetic phase (where the RS solution is exact) the distribution $P(u)$ made of $\Delta$-spaced delta peaks solves the BP equations and is stable with respect to small perturbations. What was less obvious is that starting from a generic initial condition (e.g., we start with a distribution uniform in $[-1,1]$) the population dynamics algorithm always converges to this solution in the paramagnetic phase.
In the spin glass phase the presence of the continuous part in $P(u)$ is due to the instability of the Dirac deltas with respect to any perturbation: the only compromise is the coexistence of these delta peaks with a continuous part.
We have checked that, as expected, the weight of the continuous part goes to zero at the critical point, which can be easily identified by the study of the largest eigenvalues $\mu$ of the following linear integral equation
\beq
g(u)  =  \mu M \int P_{M-1}(u_M-u_1)g(u_1)du_M \,du_1\, 
\overline{ \delta \left(u-(u_M+H)\sign\,J\right)\theta(|J|-|u_M+H|) }
\label{DATBETHET0}
\eeq 
The largest eigenvalue $\mu$ computed at $T=0$ as a function of the external field is shown in Fig.\ref{fig:lambda_h} and provides the following estimate for the critical field: $H_c(T=0) = 1.534(1)$.

At $T=0$ the eigenvector $g(u)$ presents Heaviside steps where the corresponding $P(u)$ has Dirac deltas. We show in Fig.~\ref{fig:gP_T0} the distributions $g(u)$ computed at the same field values than in Fig.~\ref{fig:Pu_T0}.
In general the distribution $g(u)$ is less singular than the corresponding $P(u)$.
We observed that $g(u)$ becomes more singular in approaching the zero temperature critical point (see Fig.~\ref{fig:Pu_Tc} and related comments below).

It is interesting to consider the large $M$ limit of the dAT line.  At finite temperature one expects to obtain the standard dAT line of the Sherrington-Kirkpatrick model. However while the dAT line of the SK model has $H_{dAT}(0)=\infty$ at zero temperature, in diluted models $H_{dAT}(0)$ is finite at any finite values of $M$ that diverges in the large $M$ limit. In order to characterize this behavior we will have to first take the $\beta \rightarrow \infty$ limit and then the $M \rightarrow \infty$ limit. The final result, derived in the appendix is:
\beq
{1 \over M^{1/2}} \simeq {2 \overline{ |J| }\over \sqrt{2 \pi \overline{J^2}}}\exp\left[-{H_{dAT}^2 \over 2 \, \overline{J^2}}\right]
\eeq
therefore $H_{dAT}$ diverges with $M$ as $H_{dAT}=\sqrt{\overline{J^2}\, \ln M}$.

\section{Six-point Susceptibilities at Criticality}
\label{cubic}

In this section we will derive expressions for the two six-point susceptibilities $\omega_1$ and $\omega_2$, whose ratio is directly related to the dynamical exponents according to eq. (\ref{uno}). We start with the computation of $\omega_1$ whose definition is: 
\beq
\omega_1={1 \over N} \sum_{ijk}\overline{\langle s_i s_j\rangle_c \langle s_j s_k\rangle_c \langle s_k s_i\rangle_c}
\eeq
We will see that $\omega_1$ diverges at criticality as $\tau^{-3}$ where $\tau$ is the same of eq. (\ref{chig}). In order to compute $\overline{\langle s_i s_j\rangle_c \langle s_j s_k\rangle_c \langle s_k s_i\rangle_c}$ we will consider only the case in which the three indices are different. Indeed one can check at the end that this is the only relevant case at criticality, because the remaining two cases give contributions that  either are not diverging or are diverging with a power less than $\tau^{-3}$. 

Let us label the spins $s_1$, $s_2$ and $s_3$ and let us call $s_0$ the spin where the three path on the tree that connects the spins $s_1,s_2$ and $s_3$ joins. This does not include the case in which, say, spin $s_1$ lies on the line connecting spin $s_2$ and $s_3$ but it can be also argued that this gives a less divergent contribution and can be neglected at the critical point.
We also call $s_{1'}$, $s_{2'}$ and $s_{3'}$ the neighbors of $s_0$ on the branches where $s_1,s_2$ and $s_3$ respectively lie.
Now let us consider the connected correlation:
\beq
\langle s_1 s_2\rangle_c={1 \over \beta} {d m_2 \over dH_1}
\eeq
Given the locally-tree-like nature of the graph the response of $m_2$ would be the same in presence of an external field on site $s_0$ proportional to the derivative of the field passed from $s_{1'}$ to $s_0$:
\beq
{d m_2 \over dH_1} \equiv {d m_2 \over dH_0}{d u_{1' \rightarrow 0} \over dH_1}
\eeq
on the other hand we have:
\beq
{d m_2 \over dH_0}={d m_0 \over dH_2}={dm_0 \over dH_0}{d u_{2' \rightarrow 0} \over dH_2}=\beta (1-m_0^2){d u_{2' \rightarrow 0} \over dH_2}
\eeq
where $m_0$ is the magnetization of site $s_0$ induced by the global cavity field acting on it:
\beq
m_0= \tanh \beta H_0 \,;\ \ \ H_0=\sum_{i \in \partial 0} u_{i \rightarrow 0} 
\eeq
Putting everything together we arrive at the following useful relationship:
\beq
\langle s_1 s_2\rangle_c=(1-m_0^2){d u_{2' \rightarrow 0} \over dH_2}{d u_{1' \rightarrow 0} \over dH_1}
\label{useful}
\eeq
Using the above relationship for $\langle s_1 s_3\rangle_c$ and $\langle s_2 s_3\rangle_c$ we finally obtain:
\beq
\langle s_1 s_2\rangle_c\langle s_2 s_3\rangle_c\langle s_3 s_1\rangle_c=(1-m_0^2)^3\left({d u_{2' \rightarrow 0} \over dH_2}\right)^2\left({d u_{1' \rightarrow 0} \over dH_1}\right)^2 \left( {d u_{3' \rightarrow 0} \over dH_3}  \right)^2
\label{con1}
\eeq
In the next step we have  to average the above expression over the disorder and the position of site $s_1$, $s_2$ and $s_3$ and the $N$ possible values of the central spin $s_0$.
It is clear that the three terms ${d u_{1' \rightarrow 0} \over dH_1}$, ${d u_{2' \rightarrow 0} \over dH_2}$ and ${d u_{3' \rightarrow 0} \over dH_3}$ are uncorrelated between each other, however they are correlated with the corresponding messages $u_{1' \rightarrow 0}$, $u_{2' \rightarrow 0}$ and $u_{3' \rightarrow 0}$. Therefore we can perform the integration over them with the help of the function $\chi(u)$ defined in eq. (\ref{guphys}). In the end we arrive at the following expression:
\beqa
{1 \over N} \sum_{i \neq j \neq k}\overline{\langle s_i s_j\rangle_c \langle s_j s_k\rangle_c \langle s_k s_i\rangle_c} & = & {M+1 \choose 3}\int du_1 du_2 du_3 du_{M-2} \, \chi(u_1)\chi(u_2)\chi(u_3)P_{M-2}(u_{M-2})\times 
\nonumber
\\
& \times & (1 - m_0^2)^3 + o(\tau^{-3})
\eeqa
where 
\beq
m_0= \tanh \beta[u_1+u_2+u_3+u_{M-2}+H]
\eeq
and $P_{M-2}(u_{M-2})$ is the distribution of the sum of $M-2$ fields distributed independently according to the function $P(u)$. According to eq. (\ref{chig}) at criticality the joint susceptibility $\chi(u)$ diverges and can be written as a solution $g(u)$ of the homogeneous equation (\ref{DATBETHE}) times a constant diverging as the inverse of the distance from the critical point $\tau$. Then it follow that $\omega_1$ diverges as $\tau^{-3}$.
We stress that the above expression is only valid at leading order and we can now show that the cases we did not consider give contributions that are less divergent at criticality.
It is immediate to verify that the case in which the three spins are equal gives a contribution that remains finite at criticality. The case in which only two spins are equal can be obtained following the derivation of section \ref{derbethe} assuming that the two coinciding spins are the located on the root, the final result is 
\beq
{1 \over N} \sum_{i \neq j}\overline{\langle s_i s_j\rangle_c^2 (1-\langle s_i\rangle^2)}=(M+1)\int P_{M}(u')\chi(u'') [1-\tanh^2(\beta H+ \beta (u'+u''))]^3\, du' du'' \ ,
\eeq
from this we see immediately that this quantity diverges only as $\tau^{-1}$ at criticality.
Finally the case in which the three spins are different but are arranged on a single path is equivalent in the above framework to the assumption that one of the three spins coincides with $s_0$ and it is straightforward to verify that this gives a contribution diverging as $\tau^{-2}$.

Now we turn to the computation of the second cumulant
\beq
\omega_2={1 \over 2 N} \sum_{ijk}\overline{\langle s_i s_j s_k\rangle_c^2 }
\label{omega22}
\eeq
We proceed as above and we write:
\beq
\langle s_1 s_2 s_3\rangle_c = {1 \over \beta^2} {d m_2 \over dH_1 dH_3}
\eeq
this can be obtained deriving equation (\ref{useful}) with respect to $H_3$. It is evident that the only term that depends on $H_3$ is the field $u_{3' \rightarrow 0}$ entering in the expression of $m_0$, therefore we can write:
\beq
\langle s_1 s_2 s_3\rangle_c=2 m_0 (1-m_0^2){d u_{1' \rightarrow 0} \over dH_1}{d u_{2' \rightarrow 0} \over dH_2}{d u_{3' \rightarrow 0} \over dH_3}
\label{con2}
\eeq
Squaring the above expression and proceeding as above we can write:
\beqa
{1 \over 2N} \sum_{i \neq j \neq k}\overline{\langle s_i s_j s_k\rangle_c^2} & = & {M+1 \choose 3}\int du_1 du_2 du_3 du_{M-2} \, \chi(u_1)\chi(u_2)\chi(u_3)P_{M-2}(u_{M-2})\times
\nonumber
\\
& \times & 2 m_0^2(1 - m_0^2)^2  + o(\tau^{-3})
\eeqa
The first term corresponds to the assumption that the three spins are different and are connected through a spin $s_0$ different from each of them.
We can easily repeat the analysis for $\omega_1$ and show that this term gives a contribution diverging as $\tau^{-3}$ at criticality while the other terms in (\ref{omega22}) give less divergent contributions.

Since in the critical region $\chi(u)$ is proportional to $g(u)$ according to eq. (\ref{chig}) we can now express the coefficient $\omega_2 / \omega_1$ as:
\beq
{\omega_2 \over \omega_1}={ \langle\langle 2 m_0^2(1 - m_0^2)^2 \rangle\rangle \over \langle\langle (1 - m_0^2)^3 \rangle\rangle}
\label{w2suw1bis}
\eeq
where 
\beq
m_0= \tanh \beta[u_1+u_2+u_3+u_{M-2}+H]
\label{u0bis}
\eeq
and
\beq
\langle\langle \cdots \rangle\rangle = \int du_1 du_2 du_3 du_{M-2} \, g(u_1)g(u_2)g(u_3)P_{M-2}(u_{M-2}) \cdots \ .
\eeq
This completes the derivation of eq. (\ref{w2suw1}), we note that in the large $M$ limit one can easily check that the above expression reduces to the results of Sompolinsky and Zippelius for the SK model, see Eqs.~(6.20) and (6.21) in \cite{Sompolinsky82}.

\begin{figure}
\includegraphics[width=0.55\columnwidth]{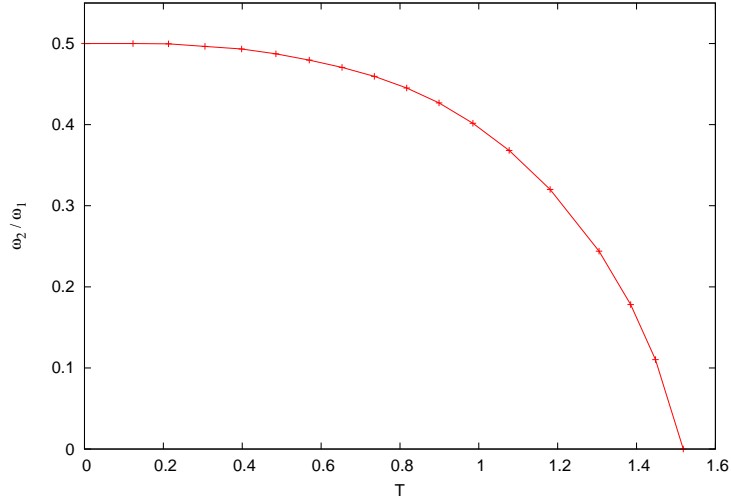}
\caption{The ratio $\omega_2/\omega_1$ computed along the critical dAT line as a function of the temperature for the Bethe lattice SG with connectivity $c=4$. The ratio tends to $1/2$ at zero temperature.}
\label{fig:omegaRatio}
\end{figure}

It is also interesting to consider the zero temperature limit of the ratio $\omega_2 \over \omega_1$. In order to do so we have to consider the distribution of the variable $u_0=u_1+u_2+u_3+u_{M-2}+H$ in (\ref{u0bis}). If this variable has a continuous distribution $P_0(u_0)$ in the $T \rightarrow 0$ limit we can make the rescaling $\beta u_0=y$ in (\ref{w2suw1bis}). Now the region relevant for the integrals is the region corresponding to $u_0=0$ where $P_0(u_0)$ can be replaced by $P(0)$, the net result is:
\beq
{\omega_2 \over \omega_1}={  \int_{-\infty}^{\infty} 2 \tanh^2\, y\,(1-\tanh^2\,y)^2 dy \over  \int_{-\infty}^{\infty} (1-\tanh^2\,y)^3 dy}={1 \over 2}
\eeq
Note that this result holds independently of the connectivity and it also coincide with the result for the SK model in the $T \rightarrow 0$ limit.

In figure (\ref{fig:omegaRatio}) we plot the ratio $\omega_2/\omega_1$ computed according to the formula (\ref{w2suw1bis}) on the dAT line of the Bethe lattice model with connectivity $M+1=4$. The data shown satisfy the expected zero-temperature limit $\omega_2/\omega_1=1/2$. The ratio increases from zero to $1/2$ upon lowering the temperature and correspondingly the dynamical exponent $a$ defined by eq. (\ref{uno}) decreases from $1/2$ to $.395$.
The value of $a=.404$ that can be red for $H=.7$ was compared in previous work with numerical data, displaying a very good agreement, see fig. 1 in \cite{calta1}.

\section{Conclusions}
\label{Conclusions}

We have presented a method, based on cavity arguments, to compute the spin-glass and higher-order susceptibilities in diluted mean-field spin-glass models.
The divergence of the spin-glass susceptibility is associated to the existence of a non-zero solution of a homogeneous linear integral equation. 
Six-point susceptibilities, relevant for the $q(x)$ function in  RSB phase and for critical dynamics through the parameter exponent $\lambda$, can be expressed at criticality as integrals involving the critical eigenvector. The numerical evaluation of the corresponding analytic expressions down to zero temperature has been discussed together with the connection with alternative numerical methods. The method was illustrated in the context of the de Almeida-Thouless line for a spin-glass on a Bethe lattice but can be generalized straightforwardly to more complex situations. The key for the derivation is eq. (\ref{useful}) from which one can express the six-point susceptibilities in terms of the joint susceptibility $\chi(u)$ which is in turn proportional to the eigenvector $g(u)$ of the homogeneous integral equation at criticality.
We note that in the case of factor graphs, corresponding to $p$-spin interactions, one has to take into account that  the node connecting the three spins in the discussion of section \ref{cubic} can be either a factor or a variable node, but it is straightforward to derive the equivalent of eq. (\ref{useful}) for a factor node.

{\em Acknowledgments.} ~~ This research has received ﬁnancial support from the European Research Council (ERC) through grant agreement No. 247328 and from the Italian Research Minister through the FIRB project No. RBFR086NN1.

\appendix

\section{The SK limit at finite and zero temperature}

In this appendix we study the dAT line analytically in the large-$M$ limit.  {\it At any finite temperature} we will recover the standard dAT line of the Sherrington-Kirkpatrick model. It is well known that the dAT line of the SK model has $H_{dAT}(0)=\infty$ at zero temperature, instead in diluted models $H_{dAT}(0)$ is finite at any finite values of $M$ but diverges in the large $M$ limit. In order to characterize this behavior we will have to first take the $\beta \rightarrow \infty$ limit and then the $M \rightarrow \infty$ limit. 

In order to reach the large-$M$ limit we must consider rescaled couplings $J/M^{1/2}$ with $J$ finite.
As a consequence the distribution $P_{M}(u_M)$ becomes a Gaussian with a finite variance. The distribution $P(u)$ instead is concentrated on very small values of $u$ and it is appropriate to consider the distribution of the variable $y=u M^{1/2}$. The distribution of $y$ is given according to eq. (\ref{PU}) by:
\beq
P(y)=\int P_M(u_M)du_M \overline{ \delta \left(y-{M^{1/2} \over \beta} \arctanh[\tanh {\beta J \over M^{1/2}} \tanh[ \beta u_M+ \beta H]]\right) }
\eeq
In the large $M$ limit we have:
\beq
{M^{1/2} \over \beta} \arctanh[\tanh {\beta J \over M^{1/2}} \tanh[ \beta u_M+ \beta H]] \rightarrow J \tanh[ \beta u_M+ \beta H]
\eeq
The function $P_M(u_M)$ according to eq. (\ref{PUM}) becomes a Gaussian in the large-$M$ limit with a variance equal to the variance of $y$, this leads to the standard replica symmetric equation of the SK model:
\beq
q=\int P(z)\tanh[ \beta z+ \beta H]]^2\, dz \ \ , \ P(z)={1 \over \sqrt{2 \pi q \, \overline{J^2}}}\exp\left[-{z^2 \over 2 q \, \overline{J^2}}\right]
\eeq
In order to write the dAT condition (\ref{DATBETHE}) in the SK limit we note that the function $g(u)$ is also concentrated around small values of $u$ and can be approximated with a delta function in the r.h.s. of eq. (\ref{DATBETHE}).  Integrating eq. (\ref{DATBETHE}) in $u$ one obtains the following homogeneous equation for $g \equiv \int g(u)du$:
\beqa
g  =  M \int P(z)\, dz \, \overline{ \left( {\sinh [{2 \beta J \over M^{1/2}}] \over \cosh [{2 \beta J \over M^{1/2}}]  + \cosh [2 \beta (z+H)]} \right)^2 } \, g
\label{DATBETHESK1}
\eeqa 
where we have used the following alternative representation of $d\tilde{u}/dh$:
\beq
\frac{\tanh(\beta J) [1-\tanh(\beta h)^2]}{[1-\tanh(\beta J)^2\tanh(\beta h)^2]} = {\sinh [2 \beta J] \over \cosh [2 \beta J]  + \cosh [2 \beta h]} 
\eeq
In the large-$M$ limit we have: 
\beq
\lim_{M\rightarrow \infty} M\left({\sinh [{2 \beta J \over M^{1/2}}] \over \cosh [{2 \beta J \over M^{1/2}}]  + \cosh [2 \beta (z+H)]}\right)^2= J^2 \beta^2 (1-\tanh^2[\beta(z+H)])^2
\eeq
thus we recover the dAT line for the SK model:
\beq
1= \int P(z)\, dz \, \overline{ J^2}  \beta^2 (1-\tanh^2[\beta(z+H)])^2
\eeq
The zero temperature limit of this equation can be obtained noticing that the variance of the Gaussian distribution $P(z)$ goes to  $\overline{J^2}$ and that 
\beq
\lim_{\beta \rightarrow \infty} \beta (1-\tanh^2[\beta z])^2 = {4 \over 3}\,\delta(z) \ .
\eeq
This leads to:
\beq
T \simeq  {4\, \overline{J^2}^{1/2}\over 3\sqrt{2 \pi}}\exp\left[-{H^2 \over 2 \overline{J^2}}\right]
\eeq
As a consequence $H_{dAT}(T)$ goes to infinity at low temperatures. On the other hand it must remain finite at any finite $M$ and in order to get its behavior we must take the large $\beta$ limit before the large $M$ limit.
In this case we can proceed as before in order to get to eq. (\ref{DATBETHESK1}), however in the next equation we have to take the $\beta \rightarrow \infty$ first and due to its non-linearity this gives:
\beq
\lim_{\beta \rightarrow \infty} M\left({\sinh [{2 \beta J \over M^{1/2}}] \over \cosh [{2 \beta J \over M^{1/2}}]  + \cosh [2 \beta (z+H)]}\right)^2= M \theta\left( {|J|\over M^{1/2}}- |z+H|\right)
\label{LMT}
\eeq
where $\theta(z)$ is the step function. Taking the $M \rightarrow \infty$ of the above equation we get:
\beq
\lim_{M \rightarrow \infty}  M \theta\left( {|J|\over M^{1/2}}- |z|\right) \approx  2 |J|M^{1/2}\delta\,(z)
\eeq
Substituting back into eq. (\ref{DATBETHESK1}) we obtain the dAT equation in the large-$M$ limit:
\beq
{1 \over M^{1/2}} \simeq {2 \overline{ |J| }\over \sqrt{2 \pi \overline{J^2}}}\exp\left[-{H^2 \over 2 \, \overline{J^2}}\right]
\eeq
therefore $H_{dAT}$ diverges with $M$ as $H_{dAT}=\sqrt{\overline{J^2}\, \ln M}$.
One may question the validity of the above result noticing that we used the Gaussian approximation for the function $P_M(u_M)$ while i) $H$ is diverging with $M$ (although logarithmically) and ii) according to eq. (\ref{LMT}) we are basically integrating it on a region of size $M^{-1/2}$ where the function does not look at all like a Gaussian (consider for instance the case $J=\pm 1$). The result however is actually correct as can be seen by means a more precise analysis including corrections that we do not report for reason of space. Such a computation can be done considering the large $M$ limit of Eq.~(\ref{DATBETHET0}) and rewriting the integral in the r.h.s. by means of a Fourier transform.

\end{document}